\newif\ifarxiv

\arxivtrue

\ifarxiv

\documentclass[letterpaper,10pt,twocolumn]{article}

\usepackage[top=1.5cm,bottom=2.0cm,left=1.5cm,right=1.5cm]{geometry}

\usepackage{sectsty}

\sectionfont{\fontsize{13}{16}\selectfont}

\else

\documentclass[letterpaper,10pt,conference]{ieeeconf}     

\IEEEoverridecommandlockouts                              

\fi 

\usepackage{times} 

\usepackage{graphicx}
\usepackage{color}
\usepackage{version}
\usepackage{layouts}
\usepackage{hyperref}
\hypersetup{
	unicode=true,
	colorlinks=true,
	citecolor=blue,
	bookmarks=true,
	breaklinks=true,
	pdfborder={0 0 0}
}
\usepackage{lcbmaths}

\newcommand\eps     {\varepsilon}
\renewcommand\phi   {\varphi}

\newcommand\beps    {\bm\eps}
\newcommand\bu      {\bm{u}}
\newcommand\bx      {\bm{x}}
\newcommand\by      {\bm{y}}
\newcommand\bz      {\bm{z}}

\newcommand\bGamma  {\bm\Gamma}
\newcommand\bLambda {\bm\Lambda}

\newcommand\hF      {\hat{F}}

\newcommand\ry      {{[y]}}
\newcommand\rh      {{\bm(h\bm)}}
\newcommand\rk      {{\bm(k\bm)}}
\newcommand\rhpo    {{\bm(h+1\bm)}}
\newcommand\rhw     {{\bm(1\bm)}}
\newcommand\rht     {{\bm(h\bm)\trop}}
\newcommand\mh      {{\bm{\{}h\bm{\}}}}
\newcommand\mht     {{\bm\{h\bm\}\trop}}
\newcommand\mhi     {{\bm\{\infty\bm\}}}
\newcommand\hist    {{-\infty:t}}
\newcommand\futt    {{t+1:\infty}}
\newcommand\ytau    {{\bm<y;\tau\bm>}}
\newcommand\abtau   {{\bm<\!\tau\!\bm>}}

\newcommand\ints    {\mathbb{Z}}

\newcommand\trop    {\mathrm{\scalebox{0.60}{\textsf{T}}}}

\newcommand\ivop    {{-1}}

\renewcommand\expect[1]  {\expectop\!\bracs{#1}}
\renewcommand{\cprob}[2]   {\prob{\rcond{#1}{#2}}}
\renewcommand{\cexpect}[2] {\expect{\rcond{#1}{#2}}}

\renewcommand\eqref[1]     {(\ref{#1})}
\newcommand\eqreff[2]      {(\ref{#1},\,\ref{#2})}
\newcommand\eqrefff[3]     {(\ref{#1},\,\ref{#2},\,\ref{#3})}
\newcommand\secref[1]      {Section~\ref{#1}}
\newcommand\figref[1]      {Fig.~\ref{#1}}
\newcommand\tabref[1]      {TABLE~\ref{#1}}

\ifarxiv

\title{\LARGE \bf
Inferring the temporal structure of directed functional connectivity in neural systems: some extensions to Granger causality}

\author{Lionel Barnett$^{*\dagger}$ and Anil K. Seth$^*$
\thanks{Sackler Centre for Consciousness Science, Department of Informatics, University of Sussex, Falmer, Brighton, BN1 9QJ, United Kingdom}%
\thanks{Corresponding author: {\tt\small l.c.barnett@sussex.ac.uk}}%
}

\else

\title{\LARGE \bf
Inferring the temporal structure of directed functional connectivity in neural systems: some extensions to Granger causality*}

\author{Lionel Barnett$^{1\dagger}$ and Anil K. Seth$^{1}$
\thanks{*This work was supported by the Dr. Mortimer and Theresa Sackler Foundation.}
\thanks{$^{1}$Sackler Centre for Consciousness Science, Department of Informatics, University of Sussex, Falmer, Brighton, BN1 9QJ, United Kingdom}%
\thanks{$^\dagger$ Corresponding author: {\tt\small l.c.barnett@sussex.ac.uk}}%
}

\fi

\begin{document}

\maketitle
\thispagestyle{empty}
\pagestyle{empty}


\begin{abstract}

Neural processes in the brain operate at a range of temporal scales. Granger causality, the most widely-used neuroscientific tool for inference of directed functional connectivity from neurophsyiological data, is traditionally deployed in the form of one-step-ahead prediction regardless of the data sampling rate, and as such yields only limited insight into the temporal structure of the underlying neural processes. We introduce Granger causality variants based on multi-step, infinite-future and single-lag prediction, which facilitate a more detailed and systematic temporal analysis of information flow in the brain.

\end{abstract}


\section{Introduction} \label{sec:intro}

Granger causality (henceforth GC) \cite{Granger:1963,Granger:1969} is a statistical, predictive notion of causal influence originally developed in econometrics, which may be inferred from time-series data, and intuitively interpreted as information flow \cite{Barnett:tegc:2009,Barnett:teml:2012}. Over the past couple of decades it has rapidly become a popular tool for the inference from neurophysiological time series data of time-directed functional (\ie, statistical) relationships in the underlying neural dynamics.

Brain recording modes such as M/EEG, ECoG and fMRI may be characterised as the discrete, regular sampling of continuous-time analogue signals associated with underlying neural processes \cite{Barnett:downsample:2017}. Due to variation in biophysical parameters such as axonal length, diameter, conduction velocity, myelination and synaptic delay \cite{Miller:1994,BuddKisvarday:2012,CaminitiEtal:2013}, such processes typically feature signal propagation delays at a range of time scales. Typical application of GC, however, involves prediction only at the time scale of a single time step into the future with respect to the chosen sampling rate. Econometricians have long known that time aggregation of signals engendered by discrete subsampling can induce spurious GC inference \cite{ComteRenault:1996,RenaultEtal:1998,BreitungSwanson:2002,McCrorieChambers:2006,Solo:2007,Solo:2016} and confound detection of actual GCs \cite{Barnett:gcfilt:2011,Seth:gcfmri:2013,Zhou:2014,Barnett:downsample:2017}.

There has, in addition, been an awareness that restriction to single-step prediction may obscure the temporal details of causal interactions within a system \cite{Hsiao:1982,Lutkepohl:1993,DufourRenault:1998}, potentially leading to misinterpretation of GC inferences. Dufour and Renault in particular \cite{DufourRenault:1998} present a thorough analysis of GC based on multiple prediction horizons, deriving algebraic conditions for (non-)causality between constituent sub-processes at all time scales. Faes~\etal\ \cite{FaesEtal:2017b,FaesEtal:2017c} present a distinct approach, where causal time scales are explored through causal filtering followed by downsampling. Here, building on \cite{DufourRenault:1998}, we present quantitative statistics for multi-step GC, and demonstrate how they may be estimated in sample and deployed for statistical inference. We also present an infinite-future GC statistic which summarises the total directed connectivity at all accessible predictive time scales, and a single-lag GC which forensically identifies individual causal feedback at specific time lags. The intention is that these statistical tools facilitate unpicking the rich multi-scale temporal details of directed functional interactions in complex neural dynamics.

\section{Wiener-Granger causality} \label{sec:gc}

Wiener-Granger causality \cite{Wiener:1956,Granger:1963,Granger:1969} is premised on a notion of causation whereby cause (i) precedes effect, and (ii) contains unique information about the effect. Formally, we suppose given a discrete-time, $n$-dimensional vector\footnote{All vector quantities are taken to be \emph{column} vectors.} stochastic process $\bu = \{\lcond{\bu_t}{t \in \ints}\}$ representing the ``universe of available information''. We introduce the notation $\bu_{t_1:t_2}$ for the range $\{\lcond{\bu_t}{t_1 \le t \le t_2}\}$, so that in particular $\bu_\hist = \{\lcond{\bu_s}{s \le t}\}$ denotes the infinite history of $\bu$ up to and including time $t$.

Suppose now that $\bu$ is partitioned into non-overlapping sub-processes, $\bu_t = [\bx_t^\trop \: \by_t^\trop \: \bz_t^\trop]^\trop$, of dimension $n_x, n_y, n_z$ respectively. We say that $\by$ does \emph{not} Granger-cause $\bx$ (at time $t$) iff
\begin{equation}
	\cprob{\bx_{t+1}}{\bu_\hist} = \cprob{\bx_{t+1}}{\bu^\ry_\hist}, \label{eq:wgc}
\end{equation}
where $\cprob\cdot\cdot$ denotes conditional distribution, and $\bu^\ry = [\bx_t^\trop \: \bz_t^\trop]^\trop$ the ``reduced'' universe of information with $\by$ omitted. Intuitively, \eqref{eq:wgc} says that removing the influence of $\by$ from the historical information set makes no difference to the statistical distribution of $\bx$ at the next time step, and we say that $\by$ Granger-causes $\bx$ iff \eqref{eq:wgc} does \emph{not} obtain. Granger \cite{Granger:1963,Granger:1969} operationalised this definition in terms of (linear) prediction:
\begin{quote}
	$\by$ Granger-causes $\bx$ iff the history of $\by$ improves prediction of the future of $\bx$ beyond the extent to which $\bx$ is already predicted by all other available historical information, including that of $\bx$ itself.
\end{quote}
Of course in practice, the ``universe of available information'' will be restricted to a specified set of accessible observables.

Granger causality was subsequently quantified by Geweke \cite{Geweke:1982,Geweke:1984} as a log-likelihood ratio statistic, and more recently \cite{Barnett:tegc:2009,Barnett:teml:2012} granted an information-theoretic (non-parametric) interpretation in terms of the closely-related \emph{transfer entropy} \cite{Schreiber:2000,Palus:2001} (in fact if all stochastic variables are jointly Gaussian \cite{Barnett:tegc:2009}, the concepts coincide). Under this interpretation---which we prefer---Granger causality represents a ``flow of information'' from the process $\by$ to the process $\bx$.

\section{Granger-Geweke causality}

Suppose now that the process $\bu$ is covariance-stationary, and without loss of generality we assume it to be zero-mean. Then by Wold's theorem \cite{Rozanov:1967,Lutkepohl:2005}, $\bu$ has a moving average (MA) representation
\begin{equation}
	\bu_t = \sum_{k = 0}^\infty B_k \beps_{t-k}\,, \text{ or }\  \bu_t = B(z) \beps_t\,, \label{eq:marep}
\end{equation}
where $\beps$ is a white noise process with nonsingular covariance matrix $\Sigma = \expect{\beps_t \beps_t^\trop}$, $z$ is the lag (backshift) operator\footnote{Note that in the literature, the lag operator is sometimes taken as $z^\ivop$. In the spectral domain, $z$ may be viewed as residing on the unit circle in the complex plane: $z = e^{-i\omega}$, where $\omega$ is the phase angle in radians.} (so that $z\cdot\beps_t = \beps_{t-1}$, \etc), and the MA operator (transfer function) is given by $B(z) = \sum_{k = 0}^\infty B_k z^k$ with $B_k$ the MA coefficient matrices, and $B_0 = I$ (the identity matrix), so that $B(z)$ is causal (does not reference the future). We also assume the minimum-phase condition that $B(z)$ is nonsingular on the closed unit disc in the complex plane, so that the MA representation \eqref{eq:marep} may be inverted to yield a stable, causal autoregressive (AR) representation
\begin{equation}
	\bu_t = \sum_{k = 1}^\infty A_k \bu_{t-k} + \beps_t\,, \text{ or }\  A(z) \bu_t = \beps_t\,, \label{eq:arrep}
\end{equation}
where $A(z) = B(z)^\ivop = I-\sum_{k = 0}^\infty A_k z^k$ is also nonsingular on the closed unit disc (see, \eg, \cite{Masani:1966,Rozanov:1967,Geweke:1982}).

Granger considered prediction in the linear least-squares sense. The optimal linear prediction of $\bu_{t+1}$ given its history $\bu_\hist$ is the conditional expectation \cite{Hamilton:1994}
\begin{equation}
	\cexpect{\bu_{t+1}}{\bu_\hist} = \sum_{k = 1}^\infty A_k \bu_{t+1-k}\,, \label{eq:upred}
\end{equation}
with residual prediction error $\beps_{t+1}$. Following \cite{Geweke:1982}, prediction error is quantified by the determinant $\Det\Sigma$ of the residuals covariance matrix, also known as the generalised variance \cite{Wilks:1932,Barrett:pre:2010}. Considering now the partition $\bu_t = [\bx_t^\trop \: \by_t^\trop \; \bz_t^\trop]^\trop$, the optimal linear prediction $\cexpect{\bx_{t+1}}{\bu_\hist}$ of $\bx_{t+1}$ given the full history $\bu_\hist$ has prediction error $\beps_{x,t+1}$ with generalised variance $\Det{\Sigma_{xx}}$ (here subscript `$x$' denotes $x$-component). We contrast this with the optimal prediction $\cexpect{\bx_{t+1}}{\bu^\ry_\hist}$ of $\bx_{t+1}$ on the reduced universe of historical information $\bu^\ry = [\bx_t^\trop \: \bz_t^\trop]^\trop$, where $\by$ is omitted; \cf, \eqref{eq:wgc} . This is derived from the reduced AR representation
\begin{equation}
	\bu^\ry_t = \sum_{k = 1}^\infty A^\ry_k \bu^\ry_{t-k} + \beps^\ry_t\,, \text{ or }\  A^\ry(z) \bu^\ry_t = \beps^\ry_t\,, \label{eq:tarrep}
\end{equation}
so that the optimal prediction of $\bu^\ry_{t+1}$ on its own history is $\cexpect{\bu^\ry_{t+1}}{\bu^\ry_\hist} = \sum_{k = 1}^\infty A^\ry_k \bu^\ry_{t+1-k}$, and the generalised variance for the optimal prediction $\cexpect{\bx_{t+1}}{\bu^\ry_\hist}$ is $\Det{\Sigma^\ry_{xx}}$ where $\Sigma^\ry = \expect{\beps^\ry_t \beps^{\ry\trop}_t}$. Following \cite{Geweke:1984} the Granger-Geweke causality statistic is defined as
\begin{equation}
	F_{\by\to\bx|\bz} = \log\frac{\Det{\Sigma^{[y]}_{xx}}}{\Det{\Sigma_{xx}}} \label{eq:ggc}
\end{equation}
and we have, in particular,
\begin{equation}
	F_{\by\to\bx|\bz} = 0 \iff A_{xy}(z) \equiv 0\,. \label{eq:ggc0}
\end{equation}

In finite sample, where the infinite histories are truncated at some model order $p$ and models \eqref{eq:arrep} and \eqref{eq:tarrep} estimated by maximum likelihood (\eg, an OLS), the estimated generalised variances are proportional to the likelihoods, and the sample estimator $\hF_{\by \to \bx | \bz}$ is a log-likelihood ratio statistic. In this scenario, \eqreff{eq:arrep}{eq:tarrep} are nested linear autoregression models, and the null hypothesis of vanishing Granger causality is
\begin{equation}
	H_0 \;:\; A_{k,xy} = 0\,, \qquad k = 1,\ldots,p\,. \label{eq:gcH0}
\end{equation}
Thus, by the standard large-sample theory \cite{Wilks:1938,Wald:1943}, under the null hypothesis \eqref{eq:gcH0} the maximum-likelihood estimator $\hF_{\by \to \bx | \bz}$, scaled by sample size, converges in distribution to a central $\chi^2(d)$ with degrees of freedom given by $d = p n_x n_y$, and a non-central $\chi^2(d;\lambda)$ with non-centrality parameter $\lambda = F_{\by \to \bx | \bz}$ under the alternative hypothesis\footnote{We remark that in sample, the scaled lack-of-fit sum of squares $[\trace{\Sigma^\ry_{xx}}-\trace{\Sigma_{xx}}]/\trace{\Sigma_{xx}}$ is asymptotically $F$-distributed under $H_0$, furnishing an alternative and more statistically powerful test for the null. However, the F-statistic lacks an information-theoretic interpretation, as well as some crucial invariance properties \cite{Barrett:pre:2010,Barnett:gcfilt:2011} of the log-likelihood ratio form, and is thus less satisfactory as a measure of \emph{magnitude} of Granger-causal effect.}. We note that $F_{\by \to \bx | \bz}$ is strictly non-negative, and thus biased in sample.

The above finite-sample analysis assumes that the AR models \eqreff{eq:arrep}{eq:tarrep} are \emph{independently} estimated. It is, however, known \cite{Chen:2006} that this may be problematic, in particular for spectral (frequency-domain) Granger-Geweke causality \cite{Geweke:1982,Geweke:1984} (which we do not address here). In fact, from the Spectral Factorisation Theorem \cite{Masani:1966} it follows that the reduced model \eqref{eq:tarrep} may be deduced from the full model \eqref{eq:arrep}, leading to more powerful and less biased GC estimators. There are several approaches to effecting this computationally, in the frequency domain \cite{Wilson:1972,DhamalaEtal:2008, Dhamala:2008a} and in the time domain \cite{Whittle:1963,Barnett:mvgc:2012}. More recently, \cite{Barnett:ssgc:2015,Solo:2016} show how this may be efficiently accomplished using state-space methods \cite{HandD:2012}\footnote{Sample statistics derived from single full-model estimation with spectral factorisation, however, fail to satisfy the requirements for the large-sample theory; in lieu of known distributions for these estimators, independent estimates of the full and reduced models or standard subsampling/surrogate methods, may be considered preferable for statistical inference.}; there are, furthermore, other compelling reasons to estimate Granger-Geweke causality via state-space rather than AR modelling \cite{Barnett:ssgc:2015,Solo:2016}.

\section{Multi-step Granger causality} \label{sec:hgc}

In the traditional approach, Granger causality is usually considered only in terms of one-step-ahead prediction [\cf, \eqreff{eq:wgc}{eq:upred}]; but see, \eg, \cite{Lutkepohl:1993,DufourRenault:1998}. However, as noted in \secref{sec:intro} the duration of a single time step will vary according to the sampling rate, and the magnitude of reported Granger-Geweke causality will depend crucially on the relationship between sampling frequency and underlying time scales of neural signal transmission \cite{Barnett:downsample:2017}. This suggests we examine more closely Granger causality based on an arbitrary future prediction horizon. A notion of (non-)causality consonant with the measure we consider was introduced in \cite{DufourRenault:1998}; here, for the first time (as far as we are aware), we quantify this notion with a Granger-Geweke statistic.

We require an expression for $\cexpect{\bu_{t+h}}{\bu_\hist}$, $h = 1,2,\ldots$; that is, optimal linear prediction at an arbitrary future prediction horizon $h$ (but note that the historical predictor set $\bu_\hist$ remains the same as for conventional $1$-step GC). In this case the $h$-step optimal prediction is more simply expressed in terms of the MA, rather than AR representation [\cf,~\eqref{eq:upred}]. We have  \cite{Hamilton:1994}
\begin{equation}
	\cexpect{\bu_{t+h}}{\bu_\hist} = \sum_{k = h}^\infty B_k \beps_{t+h-k}\,, \label{eq:hpred}
\end{equation}
with residual errors
\begin{equation}
	\beps^\rh_t = \sum_{k = 0}^{h-1} B_k \beps_{t+h-k} \label{eq:hres}
\end{equation}
[henceforth we use the round-bracket `$\rh$' to indicate a prediction horizon $h$ steps into the future]. Note that in general $\beps^\rh$ will \emph{not} be a white noise process. The residuals covariance matrix is given by
\begin{equation}
	\Sigma^\rh = \expect{\beps^\rh_t \beps^\rht_t} = \sum_{k = 0}^{h-1} B_k \Sigma B_k^\trop\,. \label{eq:hcov}
\end{equation}
Setting $B^\rh(z) = \sum_{k = 0}^{h-1} B_k z^k$ and $A^\rh(z) = B^\rh(z) A(z) = I - \sum_{k = h}^\infty A^\rh_k z^k$, we may derive the $h$-lagged AR form [\cf, \eqref{eq:arrep}]
\begin{equation}
	\bu_t = \sum_{k = h}^\infty A^\rh_k \bu_{t-k} + \beps^\rh_t\,, \text{ or }\  A^\rh(z) \bu_t = \beps^\rh_t, \label{eq:harrep}
\end{equation}
and the AR expression for the optimal $h$-step linear prediction [\cf, \eqref{eq:upred}]
\begin{equation}
	\cexpect{\bu_{t+h}}{\bu_\hist} = \sum_{k = h}^\infty A^\rh_k \bu_{t+h-k}\,. \label{eq:hupred}
\end{equation}
The $A^\rh_k$ satisfy the recursion relations \cite{DufourRenault:1998}
\begin{equation}
	 A^\rhpo_{h+k} = A^\rh_{h+k}+ A^\rh_h A_k\,, \quad h,k = 1,2,\ldots\,, \label{eq:arec}
\end{equation}
with $A^\rhw_k = A_k$.

We now define $h$-step Granger-Geweke causality by analogy with \eqref{eq:ggc} as \cite{Barnett:downsample:2017}
\begin{equation}
	F^\rh_{\by\to\bx|\bz} = \log\frac{\Det{\Sigma^{\ry\rh}_{xx}}}{\Det{\Sigma^\rh_{xx}}}\,, \label{eq:hggc}
\end{equation}
where $\Sigma^{\ry\rh} = \expect{\beps^{\ry\rh}_t \beps^{\ry\rh\trop}_t} = \sum_{k = 0}^{h-1} B^\ry_k \Sigma^\ry B^{\ry\trop}_k$, and we have [\cf,~\eqref{eq:ggc0}]
\begin{equation}
	F^\rh_{\by\to\bx|\bz} = 0 \iff A^\rh_{xy}(z) \equiv 0\,. \label{eq:hggc0}
\end{equation}
In contrast to the $1$-step case \eqref{eq:ggc0}, this condition will generally be nonlinear---specifically, a series of matrix polynomial identities of order $h$---in the AR coefficients $A_k$. In the unconditional case $\bz = \emptyset$, it may be shown \cite{Sims:1972,Caines:1976} that
\begin{equation}
	F_{\by\to\bx} = 0 \iff F_{\by\to\bx}^\rh \;\forall h > 0\,;
\end{equation}
however, in the conditional case, neither implication holds in general \cite{Lutkepohl:1993,DufourRenault:1998}. We note also from \eqrefff{eq:marep}{eq:hres}{eq:hcov}, that as $h \to \infty$, both $\Sigma^\rh_{xx}$ and $\Sigma^{\ry\rh}_{xx} \to \expect{\bx_t\bx_t^\trop}$, the covariance matrix of $\bx$ itself, implying \cite{Barnett:downsample:2017}
\begin{equation}
	\lim_{h \to \infty} F_{\by\to\bx|\bz}^\rh = 0\,. \label{eq:hggclim}
\end{equation}
Related analysis in a continuous-time scenario \cite{Barnett:downsample:2017} suggests that convergence in \eqref{eq:hggclim} is exponential.

From \eqref{eq:harrep} it follows that we again have nested ($h$-step AR) models, so that in finite sample with truncation at $p \ge h$, the null hypothesis of vanishing $h$-step Granger causality is [\cf,~\eqref{eq:gcH0}]
\begin{equation}
	H_0 \;:\; A^\rh_{k,xy} = 0\,, \qquad k = h,\ldots,p\,, \label{eq:hgcH0}
\end{equation}
and again the scaled maximum-likelihood sample estimator for $\hF_{\by\to\bx}^\rh$ will be asymptotically $\chi^2(d)$ under the null hypothesis \eqref{eq:hgcH0}, now with $d = (p-h+1)n_x n_y$. Computationally, multi-step GC may be estimated from AR or state-space models, using \eqreff{eq:hcov}{eq:hggc}. For AR modelling, the MA coefficients may be calculated recursively using
\begin{equation}
	 B_k = A_k + \sum_{\ell = 1}^{k-1} B_\ell A_{k-\ell}\,, \quad k = 2,3,\ldots\,, \label{eq:brec}
\end{equation}
with $B_1 = A_1$. For state-space modelling, calculation of the $B_k$ is even more straightforward (see \cite{Barnett:ssgc:2015}, eq.~4).

\section{Full-future Granger causality} \label{sec:ffgc}

Historically, the main emphasis of Granger causality analysis, especially in the econometrics literature, has been on statistical inference of (non-)causality. However, in light of the more recent interpretation of GC as a measure of information flow \cite{Barnett:tegc:2009,Barnett:teml:2012}, the Granger-Geweke statistic stands as an \emph{effect size}, which quantifies this information flow. This perspective seems to us particularly appropriate and intuitive with regard to functional analysis of neural systems. The conventional $1$-step prediction GC statistic, however, may be considered potentially misleading as a comparative effect size, insofar as it fails to take into account neural time scales and their interplay with sampling rate. It would thus be useful to have (in addition to the multi-step GC of \secref{sec:hgc}), a summary GC measure of the \emph{total} information flow between variables; \ie, from infinite past to infinite future. This motivates our introduction of a ``full-future'' GC measure, based on past-conditional prediction of the infinite future; that is, $\cexpect{\bu_\futt}{\bu_\hist}$.

We may calculate that the residuals covariance matrix of the prediction $\cexpect{\bx_{t+1:t+h}}{\bu_\hist}$ of the future of $\bx$ up to horizon $t+h$ from the full process history $\bu_\hist$, is given by the ($h \times h$)-block matrix
\begin{equation}
	\Sigma^\mh_{xx} = \bracs{\Sigma^{p,q}}_{xx}, \qquad p, q = 0,\ldots,h-1
\end{equation}
[note: we use curly braces $\bm\{h\bm\}$ to distinguish the full-future prediction horizon from the multi-step horizon $\bm(h\bm)$], where
\begin{equation}
	\Sigma^{p,q} = \sum_{k=0}^{h-1} \sum_{\ell=0}^{h-1} \delta_{p-k,q-\ell} B_k \Sigma B_\ell^\trop\,.
\end{equation}
This may be written
\begin{equation}
	\Sigma^\mh_{xx} = B^\mh_x \Sigma^{\otimes h} B^\mht_x\,,
\end{equation}
with
\begin{multline}
	B^\mh_x = \\
	\begin{bmatrix}
		B_{0,xu}    & 0   & 0 &   \cdots & 0 \\
		B_{1,xu} & B_{0,xu}   & 0 &   \cdots & 0 \\
		B_{2,xu} & B_{1,xu} & B_{0,xu}   & \cdots & 0 \\
		\vdots & \vdots & \vdots & \ddots & \vdots \\
		B_{h-1,xu} & B_{h-2,xu} & B_{h-3,xu} & \cdots & B_{0,xu}
	\end{bmatrix}
\end{multline}
where the index $u$ denotes all components $\{x,y,z\}$, and
\begin{equation}
	\Sigma^{\otimes h} =
	\begin{bmatrix}
		\Sigma   & 0  &   \cdots & 0 \\
		0 & \Sigma &   \cdots & 0 \\
		\vdots & \vdots & \ddots & \vdots \\
		0 & 0 & \cdots & \Sigma
	\end{bmatrix}
\end{equation}
is ($h \times h$)-block-diagonal.

For the reduced prediction $\cexpect{\bx_{t+1:t+h}}{\bu^\ry_\hist}$, we obtain $\Sigma^{\ry\mh}_{xx}$ the same way, replacing $\Sigma$ with $\Sigma^\ry$ and $B_k$ with $B^\ry_k$, and we define the full-future Granger causality as
\begin{equation}
	F^\mhi_{\by\to\bx|\bz} = \lim_{h\to\infty} F^\mh_{\by\to\bx|\bz}\,, \label{eq:ffggc}
\end{equation}
where
\begin{equation}
	F^\mh_{\by\to\bx|\bz} = \log\frac{\Det{\Sigma^{\ry\mh}_{xx}}}{\Det{\Sigma^\mh_{xx}}}\,. \label{eq:fggc}
\end{equation}
We conjecture that under appropriate conditions (\cf, \secref{sec:gc}) the limit in \eqref{eq:fggc} exists; this has been verified by extensive simulation.

In sample with finite AR model order, the null hypothesis for vanishing $F^\mhi{\by\to\bx|\bz}$ is identical to the null \eqref{eq:gcH0} for $1$-step Granger causality \eqref{eq:ggc}. This follows from the recursion relations \eqref{eq:arec} and expanding out the prediction $\cexpect{\bx_{t+1:t+h}}{\bu_\hist}$. Thus the statistic is not useful in its own right for statistical inference, and should rather be considered an informative quantitative measure of total past~$\,\to\,$~future information flow between two variables.

\begin{figure}
	\begin{center}
	\includegraphics{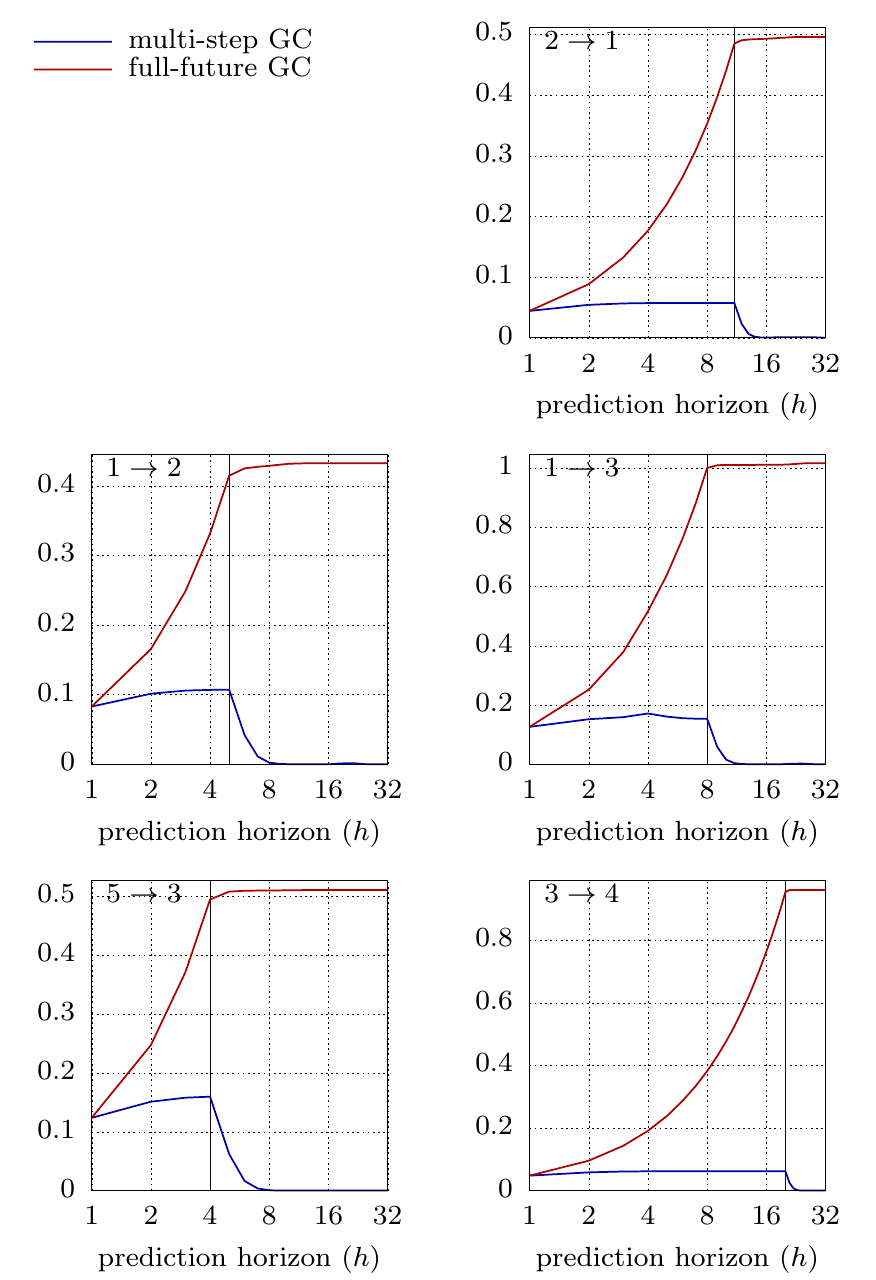}
	\end{center}
	\caption{Multi-step GC $F^\rh_{y \to x|\bz}$ (blue lines) and full-future GC $F^\mh_{y \to x|\bz}$ (blue lines) for pairs of variables $x,y$, plotted against prediction horizon $h$ (log-scale) for pairs of Granger-causal variables in a simple AR model with lagged causal feedback (see \tabref{tab:ardemo} and text for details).} \label{fig:fig_msffgc}
\end{figure}

We have not found a closed formula for the determinants in \eqref{eq:fggc}, but they may be approximated numerically; extensive simulations suggest that, although the size of the matrices $B^\mh_x$ scale quadratically in $h$, convergence to the limit in \eqref{eq:fggc} is again exponential (\cf, \figref{fig:fig_msffgc}). We remark that in general, $F^\mhi_{\by\to\bx|\bz}$ will \emph{not} be equal to the sum $\sum_{h = 1}^\infty F^\rh_{\by\to\bx|\bz}$ of multi-step Granger causalities, since the residuals $\beps^\rh_t$ of the latter \eqref{eq:hres} for different $h$ will in general be correlated, so that $\Det{\Sigma^\mh} \ne \prod_{k = 1}^h \Det{\Sigma^\rk}$.

\begin{figure}
	\begin{center}
	\includegraphics[width=0.3\textwidth]{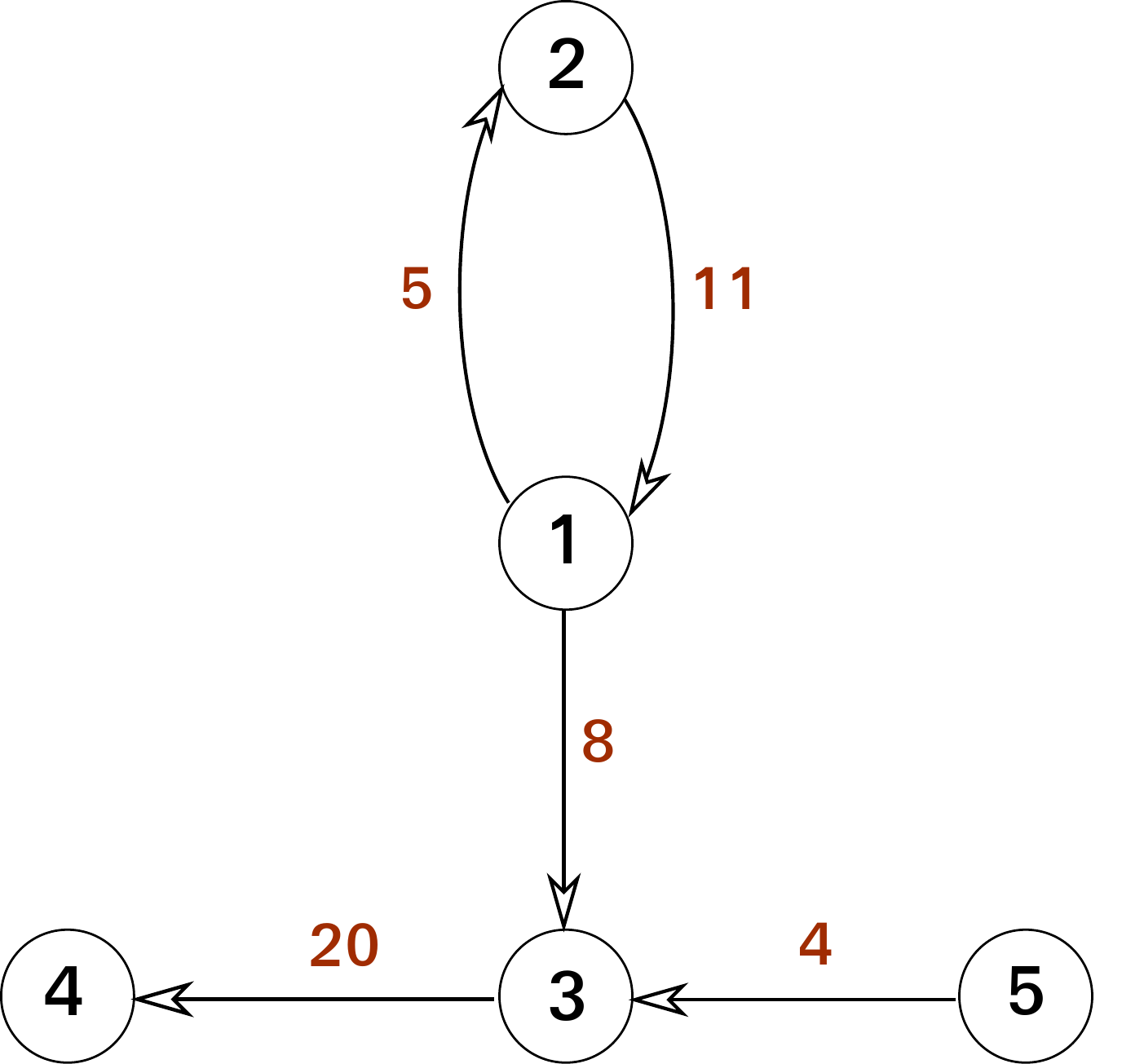}
	\end{center}
	\caption{Causal structure of the simple $5$-variable AR model (\secref{sec:ffgc}). Numbers in red denote AR lags; see \tabref{tab:ardemo} and text for details.} \label{fig:fig_model}
\end{figure}

We demonstrate multi-step and full-future GC with a simple AR model with $n=5$ variables (\figref{fig:fig_model}),
and model order $p=20$. All AR coefficients were zero except for the lag-$1$ self-regression terms $A_{1,ii}$, and the lagged coefficients set out in \tabref{tab:ardemo}.
\begin{table}
	\begin{center}
	{\renewcommand{\arraystretch}{1.2}
	\setlength{\tabcolsep}{12pt}
	\begin{tabular}{ccrr}
	\hline
		target ($x$) & source ($y$) & AR lag & AR coefficient \\
		\hline
		$1$ & $2$ & $11$ & $ 0.221 $ \\
		$2$ & $1$ & $ 5$ & $ 0.306 $ \\
		$3$ & $1$ & $ 8$ & $-0.403 $ \\
		$4$ & $3$ & $20$ & $-0.215 $ \\
		$3$ & $5$ & $ 4$ & $ 0.352 $ \\
		\hline
	\end{tabular}
	} 
	\end{center}
	\caption{Simple $5$-variable AR model parameters (\secref{sec:ffgc}).} \label{tab:ardemo}
\end{table}
The only non-zero multi-step and full-future GCs are plotted in \figref{fig:fig_msffgc}, calculated according to \eqref{eq:hggc} and \eqref{eq:fggc} respectively for prediction horizon $h = 1,\ldots,32$, with, for each directed pair of variables $x,y$, full conditioning on all remaining variables $\bz$. Note that the both the multi-step and full-future GCs coincide with the conventional $1$-step GC \eqref{eq:ggc} at prediction horizon $h = 1$. Vertical grey lines indicate the AR lag of the causal interaction (\tabref{tab:ardemo}). We see that (\cf, \cite{Barnett:downsample:2017}) the $F^\rh_{y \to x|\bz}$ decay rapidly to zero beyond the causal horizon, while the $F^\mh_{y \to x|\bz}$ rise and then quickly plateau beyond the causal horizon to the limiting value $F^\mhi_{y \to x|\bz}$ \eqref{eq:ffggc}.

\section{Single-lag Granger causality} \label{sec:slgc}

A more fine-grained analysis of directed functional connectivity may be interrogated as follows: if a variable $\by$ Granger-causes the variable $\bx$, at which \emph{specific} time scale(s) is causal feedback concentrated? Note that multi-step GC (\secref{sec:hgc}) does not directly address this question, since there all lags of a predictive (source) variable are considered together. Rather, for a specific lag $\tau > 0$, we consider a Granger statistic based on the null hypothesis
\begin{equation}
	H_0 \;:\; A_{\tau,xy} = 0\,, \label{eq:sgcH0}
\end{equation}
where $A_k$ are the AR coefficient matrices in \eqref{eq:arrep}. We thus compare the optimum prediction  $\cexpect{\bx_t}{\bu_{-\infty:t-1}}$ with the optimum prediction $\expect{\rcond{\bx_t}{\bu^\ytau_{-\infty:t-1}}}$, where the superscript `$\bm<\!y;\tau\!\bm>$' indicates that the single lag $\by_{t-\tau}$ of $\by$ is omitted from the historical predictor set $\bu_{-\infty:t-1}$. To make this clearer, consider the $x$-component of the AR representation \eqref{eq:arrep} of $\bu_t$:
\begin{align}
	\bx_t &= A_{1,xx} \bx_{t-1} + A_{2,xx} \bx_{t-2} + \ldots \notag \\
	      &+ A_{1,xx} \by_{t-1} + A_{2,xx} \by_{t-2} + \ldots + \boxed{A_{\tau,xy} \by_{t-\tau}} + \ldots \notag \\
	      &+ A_{1,xz} \bz_{t-1} + A_{2,xz} \bz_{t-2} + \beps_{xt}\,. \label{eq:slar}
\end{align}
The reduced AR representation then omits the boxed lag-$\tau$ $\by$ regressor, and we define the single-lag Granger causality as
\begin{equation}
	F^\abtau_{\by\to\bx|\bz} = \log\frac{\Det{\Sigma^\ytau_{xx}}}{\Det{\Sigma_{xx}}}\,, \label{eq:sggc}
\end{equation}
where $\Sigma^\ytau_{xx}$ is the residuals covariance matrix for the reduced AR model. We note that $F^\abtau_{\by\to\bx|\bz} = 0 \;\forall \tau > 0 \iff F_{\by\to\bx|\bz} = 0$

The regression \eqref{eq:slar} with the null condition \eqref{eq:sgcH0} represents a nested linear model, so that the large-sample theory applies, and the scaled sample estimator $\hF^{<\!\tau\!>}_{\by\to\bx|\bz}$ will thus be asymptotically $\chi^2(d)$ with $d = n_x n_y$. We remark that interpretation of $F^\abtau_{\by\to\bx|\bz}$ as an effect size for a putative ``information flow'' is somewhat moot; we may prefer to consider $F^\abtau_{\by\to\bx|\bz}$ simply as a test statistic for inference of (the absence of) a causal feedback from source to target variable at the given lag.

Unlike the previous GC measures, we do not have (given full-model parameters) a construction for a state-space model which represents the reduced model \eqref{eq:sgcH0}. The reduced model parameters may, however,  still be solved computationally from the Yule-Walker equations \cite{Lutkepohl:1993}. The full-model Yule-Walker equations up to lag $q$ yield
\begin{equation}
	\Sigma = \Gamma_0 - \bGamma_q \bLambda_q^\ivop \bGamma_q^\trop\,, \label{eq:yw}
\end{equation}
with $\Gamma_k = \expect{\bu_t \bu_{t-k}^\trop}$, $k = \ldots,-2,-1,0,1,2,\ldots$ the autocovariance sequence---which may itself be derived from the (estimated) full-model AR coefficients \cite{Barnett:mvgc:2012}---and
\begin{align}
	\bGamma_q &= \begin{bmatrix} \Gamma_1 & \cdots & \Gamma_q \end{bmatrix} \\[1em]
	\bLambda_q &=
	\begin{bmatrix}
		\Gamma_0 & \cdots & \Gamma_{q-1} \\
		\vdots   & \ddots & \vdots \\
		\Gamma_{q-1}^\trop & \cdots & \Gamma_0
	\end{bmatrix}.
\end{align}
The reduced Yule-Walker solution for $\Sigma^\ytau$ is then obtained as per \eqref{eq:yw}, after deleting the $y$-columns of the $\tau$-th block-column in $\bGamma_q$, and the $y$-rows/columns of the $(\tau-1)$-th block-row/column in $\bLambda_q$. Even though $\bLambda_q$ may be quite large\footnote{For reasonable numerical precision we need sufficient lags $q$ that $\Gamma_k \approx 0$ for $k > q$, which will in turn depend on the spectral radius of $A(z)$ \cite{Lax:2007}; see \eg, \cite{Barnett:mvgc:2012}.}, it is positive-definite Toeplitz and thus may be Cholesky-decomposed and efficiently inverted.

\begin{figure*}
	\begin{center}
	\includegraphics{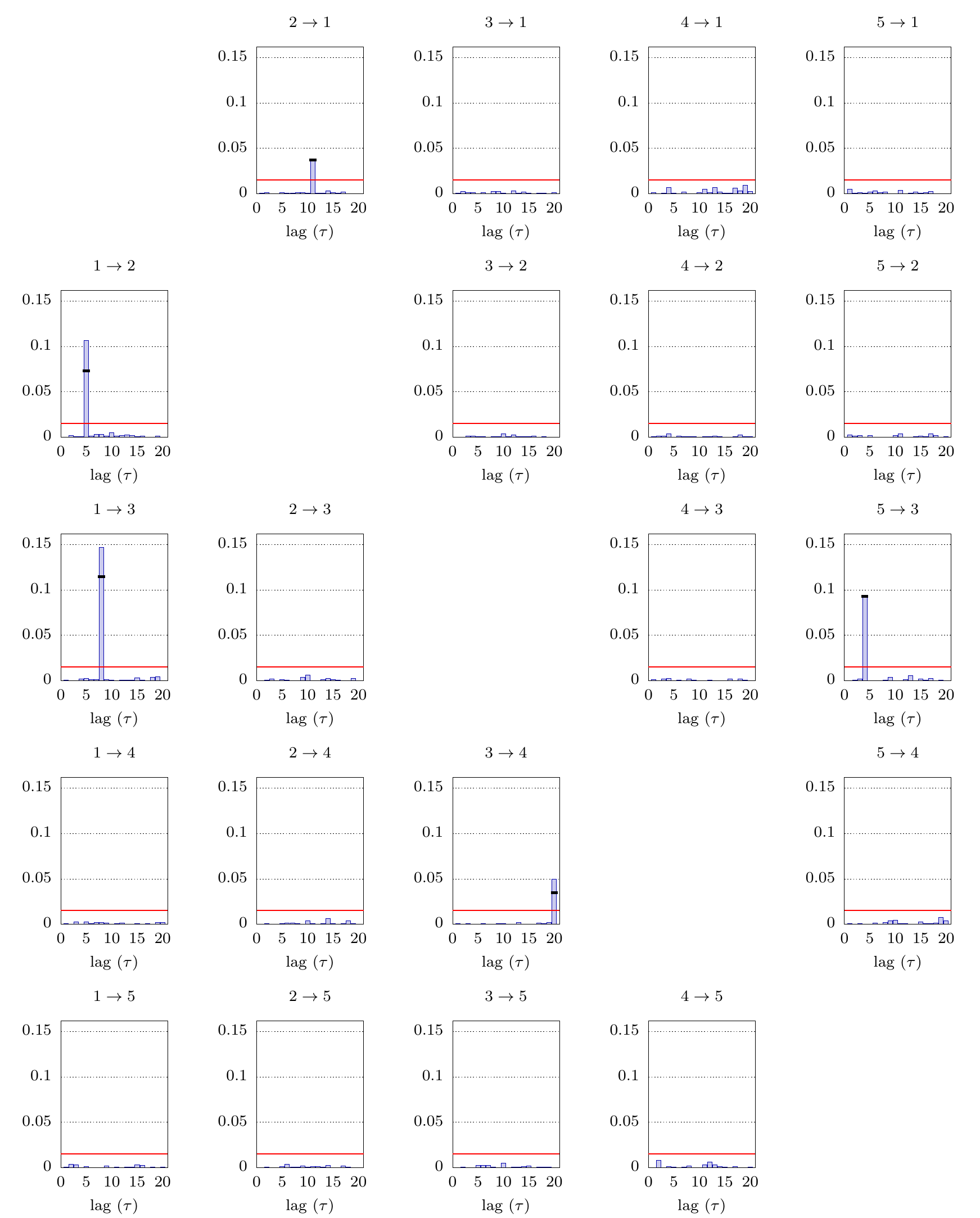}
	\end{center}
	\caption{Single-lag GC inference (``Granger-causal graph'') for time-series data generated from the simple $5$-variable AR model with varying causal lags of \secref{sec:ffgc} (\figref{fig:fig_model} and \tabref{tab:ardemo}). Blue boxes represent estimates of the single-lag GCs $F^\abtau_{y \to x|\bz}$ \eqref{eq:sggc}, while bold black horizontal bars denote actual values computed analytically. Red horizontal lines mark the critical GC level; see text (\secref{sec:slgc}) for details.} \label{fig:fig_slgc_ex}
\end{figure*}

We envisage estimating $F^\abtau_{\by\to\bx|\bz}$ from the data for $\tau = 1,\ldots,p$ in turn (where the maximum lag $p$---the model order for the full AR model \eqref{eq:slar}---is selected via a standard scheme), in order to ascertain the time scale(s) at which $\by$ influences $\bx$. See \figref{fig:fig_slgc_ex}, where the  $F^\abtau_{y \to x|\bz}$, $x,y = 1,\ldots,n$, $x \ne y$, are estimated in sample for a data sequence of length $1000$ generated from the AR model of \secref{sec:ffgc}.
Here $\bz$ denotes all other variables except the given $x,y$, so that every directed pairwise GC is conditioned on all remaining variables, yielding the ``Granger-causal graph'' \cite{Barnett:mvgc:2012} at all lags up to $p = 20$. Likelihood-ratio single-lag GC statistics (blue boxes) were calculated for separate OLS estimates of the full and (for each $j,\tau$) reduced models \eqref{eq:slar} using the (known) model order $p = 20$, while analytic GCs for the model (black horizontal bars) were calculated from the actual model parameters (\tabref{tab:ardemo}) using the Yule-Walker procedure described above with $q = 175$ autocovariance lags, which was sufficient to ensure that the $\Gamma_k$ decay to near-machine precision. The red horizontal lines mark the critical GC level for rejection of the null hypotheses \eqref{eq:sgcH0} of zero single-lag GC at significance $\alpha = 0.05$ according to the $\chi^2(1)$ estimator distribution, assuming a Bonferroni correction for all $pn(n-1)$ hypotheses. We see that statistical inference of the $F^{<\!\tau\!>}_{y \to x|\bz}$ correctly identifies the causal lags as well as directed functional connectivity in the model (\figref{fig:fig_model}).

\ifarxiv
\section{Conclusions}
\else
\section{CONCLUSIONS}
\fi

In this article we address the issue of how, in an empirical scenario, we may go beyond Granger-causal inference restricted to the time scale prescribed by the data sampling rate, to obtain a more detailed picture of Granger-causal interactions at multiple times scales underlying the measured neurophysiological process. Thus our multi-step measure $F^\rh_{\by\to\bx|\bz}$ reflects information flow between variables at a specific future time horizon $h$, while the single-lag measure $F^\abtau_{\by\to\bx|\bz}$ identifies the precise time lag(s) at which a specific Granger-causal interaction operates. Via a simple didactic model, we demonstrate, respectively, how the underlying time scales are reflected via the multi-step statistic (\figref{fig:fig_msffgc}), and may be explicitly inferred from the data (\figref{fig:fig_slgc_ex}). In addition, in our full-future GC measure $F^\mhi_{\by\to\bx|\bz}$, we present a useful summary measure of effect size for the total past~$\to$~future information flow between variables. All measures are fully conditioned on (accessible) exogenous variables, so that only \emph{direct} functional relationships are reported. We describe how our measures may be estimated computationally from time-series data, and (where appropriate) their asymptotic sampling distributions. We propose these measures as useful additions to the directed functional analysis toolbox, insofar as they stand to elucidate time-dependant causal interactions in neurophysiological processes of interest to neuroscientific research. We encourage future research into the behaviour of our measures for more realistic data, where we should expect causal interactions at multiple distributed lags \cite{Barnett:downsample:2017}.

\ifarxiv
\section*{Acknowledgments}
We are grateful to the Dr. Mortimer and Theresa Sackler Foundation, which supports the Sackler Centre for Consciousness Science.
The authors would also like to thank Stefan Haufe for useful discussions.
\else
\section*{ACKNOWLEDGMENT}
The authors would like to thank Stefan Haufe for useful discussions about this work.
\fi



\ifarxiv
\bibliographystyle{plain}
\else
\bibliographystyle{ieeepes} 
\fi

\bibliography{Barnett_SMC_2019_SS}

\end{document}